\documentclass[conference]{IEEEtran}
\usepackage{amsmath}
\usepackage{amssymb}
\usepackage{amsfonts}
\usepackage{graphicx}
\usepackage{enumerate,color,graphicx,fancybox,pifont,epsf,epsfig,subfigure,amsmath,amssymb,psfrag}
\usepackage{algorithm}
\usepackage{algorithmic}
\usepackage{cite}

\newtheorem{theorem}{\textbf{Theorem}}

\newtheorem{definition}[theorem]{\textbf{Definition}}

\newcommand{\figref}[1]{Figure~\ref{#1}}
\newcommand{\defref}[1]{Definition~\ref{#1}}

\newcommand{\theoref}[1]{Theorem~\ref{#1}}

\newcommand{\lemref}[1]{Lemma~\ref{#1}}

\newcommand{\cororef}[1]{Corollary~\ref{#1}}

\newtheorem{theo}{\textbf{Theorem}}
\newtheorem{lem}{\textbf{Lemma}}

\newtheorem{coro}{\textbf{Corollary}}

\newcommand{\qed}{\nobreak \ifvmode \relax \else
      \ifdim\lastskip<1.5em \hskip-\lastskip
      \hskip1.5em plus0em minus0.5em \fi \nobreak
      \vrule height0.75em width0.5em depth0.25em\fi}

\title{Performance Bounds for Vector Quantized Compressive Sensing}
\author{Amirpasha Shirazinia, Saikat Chatterjee, Mikael Skoglund\\
ACCESS Linnaeus Centre, KTH Royal Institute of Technology, Stockholm, Sweden\\
\texttt{Email: amishi@ee.kth.se, sach@kth.se, skoglund@ee.kth.se}
}

\begin{document}

\maketitle

\begin{abstract}
In this paper, we endeavor for predicting the performance of quantized compressive sensing under the use of sparse reconstruction estimators. We assume that a high rate vector quantizer is used to encode the noisy compressive sensing measurement vector. Exploiting a block sparse source model, we use Gaussian mixture density for modeling the distribution of the source. This allows us to formulate an optimal rate allocation problem for the vector quantizer. Considering noisy CS quantized measurements, we analyze upper- and lower-bounds on reconstruction error performance guarantee of two estimators - convex relaxation based basis pursuit de-noising estimator and an oracle-assisted least-squares estimator.

\end{abstract}
\section{Introduction} \label{sec:intro}
\vspace{-0.2cm}
Using under-determined linear set of equations, compressive sensing (CS) \cite{08:Candes} aims to reconstruct a high dimensional sparse source signal from an under-sampled low dimensional measurement vector. In the CS literature, several algorithms (see e.g., \cite{06:Candes2,07:Tropp,07:Candes,08:Shihao,09:Dai}) have been developed to recover a sparse source from the CS measurements. A number of algorithms have been also analyzed for their performance guarantees in noiseless and noisy measurement cases.

For practical applications, a CS measurement vector needs to be quantized and transmitted possibly over a noisy communication channel. CS with quantized measurements has recently started to gain significant attention in literature. Signal recovery from noisy measurements -- which can be thought of as the effect of quantization when the quantization error is known, bounded and additive -- has been addressed in \cite{06:Candes2}. In \cite{10:Zymnis}, the authors focus on convex-based recovery of a sparse signal from a set of quantized measurements. The goal of \cite{11:Dai} is to find average distortion bounds caused by quantization of CS measurements, and by using some practical recovery algorithms. The same aim has been tackled in \cite{11:Laska} where the authors explore a trade-off between number of measurements and quantization rate. In \cite{11:Jacques}, a reconstruction scheme of sparse signals from quantized measurements has been proposed based on non-Gaussianity of quantization error.

In this paper, we consider a set-up where CS measurement vector is corrupted by additive white Gaussian noise (AWGN). The CS measurement vector is assumed to be quantized using a high rate vector quantizer (VQ), reconstructed and then transmitted over the AWGN channel. We use a block sparse source signal model such that the probability density function (pdf) of the source can be efficiently modeled by a Gaussian mixture (GM) density. Applying block sparse signal model \cite{09:Stojnic,10:Stojnic,10:Eldar,10:Baraniuk} does not incur any significant loss of generality. Particularly, if we treat the dimension of a block as one, then a block sparse source becomes an unconstrained sparse source. The main motivation of using block sparse signal model along with Gaussian mixture model (GMM) is the analytical tractability. First, by employing the GMM, the noisy CS measurement vector also becomes GM distributed because of the linear relation. Second, exploiting the GMM allows ease of analytical tractability for further use of high rate theory of VQ \cite{00:Hedelin,03:Subramaniam,08:Saikat1,08:Saikat2}. By designing an optimal rate allocation for the Gaussian components of the GMM measurement vector, we guarantee an optimum VQ with respect to minimizing quantization distortion, and under this circumstance, we provide the recovery performance guarantees of two estimators - convex relaxation based basis pursuit de-noising (BPDN) algorithm and an oracle-assisted least-squares estimator. In our analysis, the quantization noise is non-Gaussian, and bounds on the performance guarantees are derived by analyzing the tail distribution of the total noise (quantization noise and the Gaussian noise sources). Using a model of quantization error, practical simulation results are provided vis-a-vis the bounds.

\emph{Notations :} Bold-faced upper case (lower case) characters are used for matrices (vectors). We also denote transpose of a matrix by $(\cdot)^T$. Further, diagonal elements of a matrix are denoted by \textbf{diag}$(\cdot)$ and determinant of a matrix by $|\cdot|$. An all-one vector of size $N$ is denoted by $\mathbf{1}_N$.  We use~$\mathbb{E}[\cdot]$ to denote the expectation operator. The cardinality of a set is identified by \textbf{card}$(\cdot)$. We represent $\ell_2$-norm of a vector by $\|\cdot\|_2$.
\vspace{-0.2cm}
\section{Problem statement} \label{sec:problem}
\vspace{-0.1cm}
We consider a random block sparse source $\mathbf{x}$ is sensed by noisy sensors that yield an under-determined set of linear transformation corrupted by an additive white noise. The noisy measurements are quantized and reconstructed using a VQ, and then these samples are conveyed over an AWGN channel. In fact, quantization is essential in practical applications such as sensor networks where transmission and computational resources are limited. In our study, let us denote the sensing matrix, measurement noise, channel noise and the quantization function, respectively, by $\mathbf{A}$, $\mathbf{n_m}$, $\mathbf{n_c}$ and $\mathcal{Q}$, where we assume that the dimension of $\mathbf{y}$ is much less than that of $\mathbf{x}$.  Then, $\mathbf{y_c} = \mathcal{Q}(\mathbf{Ax} + \mathbf{n_m}) + \mathbf{n_c}$, where $\mathbf{y_c}$ represents the channel output. At the receiver side, an estimator takes $\mathbf{y_c}$ as an input, and produces an estimate of the source, denoted by $\mathbf{\hat{x}}$. 

The performance of the system is measured by the estimation error $\mathbf{\|\mathbf{x - \hat{x}}\|_2}$. We establish the goal of our work to design the quantizer by minimizing the overall distortion caused in the system so as to guarantee robust recovery of a block sparse source from an under-determined set of noisy linear measurements. In the following sub-sections, we describe the functionality of each building block of the system. 
\vspace{-0.3cm}
\subsection{Block sparse source} \label{subsec:block sparsity}
\vspace{-0.1cm}
We apply a GMM-based block sparse source which is first introduced in~\cite{11:Mikko}. Let $M=QR$ be the length of a sparse vector $\mathbf{x}$. Further, let $\mathbf{x}$ be comprised of $R$ non-overlapping equal length sub-vectors $\{\mathbf{x}_r \}_{r=1}^R$, i.e.,
\begin{equation} \label{eq:def x}
\mathbf{x} \!=\!  \left[\mathbf{x}_1, \ldots, \mathbf{x}_R \right]^T \!\in\! \mathbb{R}^M , \textrm{where  } \mathbf{x}_r \!=\!\left[x_{r,1}, \ldots, x_{r,Q}\right]^T \!\in\! \mathbb{R}^Q.
\end{equation}
We let $\mathbf{x}$ be drawn according to the GM density as
\begin{equation} \label{eq:pdf x}
    f(\mathbf{x}) = \sum_{k=1}^K \sum_{l=1}^{L_k}  \omega_{k,l} \mathcal{N}(\mathbf{0}, \mathbf{C}_{k,l}),
\end{equation}
where $K$ (block sparsity indicator) is an integer,
\begin{equation} \label{eq:param 1}
    L_k = {{R}\choose{k}}, \hspace{0.2cm} \omega_k=\sum_{l=1}^{L_k} \omega_{k,l} \hspace{0.2cm} \textrm{and} \hspace{0.2cm}\sum_{k=1}^K \omega_k=1,
\end{equation}
for $\omega_{k,l} \!\geq \!0$. Here, $\omega_{k}$ denotes the probability of observing $k$ information bearing blocks in a realization of $\mathbf{x}$, $L_k$ the number of different arrangements of these blocks in $\mathbf{x}$, and $\omega_{k,l}$ the probabilities related to these arrangements. We let the diagonal covariance matrices in \eqref{eq:pdf x},
\begin{equation} \label{eq:cov matrix x}
    \mathbf{C}_{k,l} = \textbf{diag} \left(\mathbf{c}_1^{(k,l)},\ldots,\mathbf{c}_R^{(k,l)} \right), \hspace{0.2cm} l=1,\ldots,L_k,
\end{equation}
be different for all $(k,l) \neq (k',l')$. The block diagonal matrices $\{\mathbf{c}_r^{(k,l)}\}$ take only two values
\begin{equation} \label{eq:values cov}
    \{\mathbf{c}_r^{(k,l)}\} = \mathbf{1}_Q \theta^2 \hspace{0.2cm} \textrm{and} \hspace{0.2cm} \{\mathbf{c}_r^{(k,l)}\} = \mathbf{1}_Q \rho_{x}^2, \hspace{0.2cm} \forall{r,k,l},
\end{equation}
where the parameters $\theta^2$ and $\rho_x^2$ represent the expected signal powers of sparsity including and information bearing components of $\mathbf{x}$, respectively. 
\begin{definition} \cite[Definition 1]{11:Mikko} \label{def1}
Let $\mathbf{x}$ be a random variable (RV) with pdf \eqref{eq:pdf x}. We say $\mathbf{x}$ is an \textit{approximately} $K$ block sparse RV if $\theta^2 \!\ll\! \rho_x^2$ and all covariance matrices $\{\mathbf{C}_{k,l}\}$ satisfy \eqref{eq:cov matrix x} and \eqref{eq:values cov}. If $\theta^2 \!\to\! 0^+$, we simply say that $\mathbf{x}$ is a $K$ block sparse RV.
\end{definition}

To clarify the block sparsity $K$ over all the blocks of $\mathbf{x}$ in a set $\mathcal{R}\!=\! \{Q_1\!=Q\!,\ldots,Q_R\!=Q\!\}$, where $Q_r$'s are the equal block lengths, we denote
\begin{equation} \label{eq:block sparsity}
\|\mathbf{x}\|_{0,\mathcal{R}} \triangleq \sum_{i=1}^R \Xi_{\mathcal{R}}(\|\mathbf{x}_i\|_2>\gamma),
\end{equation}
where $\Xi_{\mathcal{R}}(\cdot)$ is an indicator function that picks $1$ if $\|\mathbf{x}_i\|_2>\gamma$, and $0$ otherwise over the set $\mathcal{R}$. Assuming $\gamma >0$ is an arbitrary small number, the block sparsity satisfies $\|\mathbf{x}\|_{0,\mathcal{R}} \leq K$. Furthermore, in the limit case when $\theta^2 \rightarrow 0^+$ then $\gamma \rightarrow 0^+$ which coincides with \cite[Definition 1]{09:Eldar}.

We introduce the sensing matrix ~$\mathbf{A}\! \in \!\mathbb{R}^{N\!\times\! M}$ ($N \ll M$), and the noisy measurement vector as~$\mathbf{y}\!=\!\mathbf{Ax \!+\! n_m} \!\in\! \mathbb{R}^{N}$, where $\mathbf{n_m} \!\sim \! \mathcal{N}(\mathbf{0},\sigma_m^2 \mathbf{I}_N)$ is independent of the linear transformation with \textit{a priori} known variance $\sigma_m^2$. In order to ensure the recovery of the vector $\mathbf{x}$ from the linear observations using reconstruction algorithms, several conditions have been deployed. One of these sufficient conditions regarding the sensing matrix is the so-called restricted isometry property (RIP)\cite{05:Candes,06:Candes} that characterizes matrices which are nearly orthogonal. Similar to the definition of conventional RIP\cite[Definition 1.1]{05:Candes}, block-RIP for recovery of block sparse signals reads as follows.
\begin{definition}\cite[Definition 2]{09:Eldar} \label{def2}
    $\mathbf{A} \!\in \!\mathbb{R}^{N \!\times\! M}$ with $\ell_2$-normalized columns is said to have the block-RIP over all the blocks of $\mathbf{x}$ in a set $\mathcal{R}$ if for a $K$-block sparse $\mathbf{x}$,
    \begin{equation} \label{eq:BRIP}
        (1-\delta_{K|\mathcal{R}})\|\mathbf{x}\|_2^2 \leq \|\mathbf{Ax} \|_2^2 \leq (1+\delta_{K|\mathcal{R}})\|\mathbf{x}\|_2^2,
    \end{equation}
where the block-RIP parameter $\delta_{K|\mathcal{R}} \! \in \!(0,1)$ is the smallest quantity satisfying \eqref{eq:BRIP}.
\end{definition}

\vspace{-0.2cm}
\subsection{GMM-based VQ} \label{VQ}
\vspace{-0.1cm}
A VQ of size $I$ is a mapping from a vector $\mathbf{y}$ in $\mathbb{R}^N$ into a discrete finite set $\mathcal{G}$ in $\mathbb{R}^N$ containing $I$ code-vectors, i.e., $\mathcal{Q}: \mathbb{R}^N \mapsto \mathcal{G}$, where the set $\mathcal{G}$ is called a code-book. The rate of the VQ is measured as $b_{t}=\log_2 I$ in bits/vector. 

In our scenario, it can be shown that the noisy measurement vector $\mathbf{y}$ has the GM density as
\begin{equation} \label{eq:pdf y}
    f(\mathbf{y}) = \sum_{k=1}^K \sum_{l=1}^{L_k}  \omega_{k,l} \mathcal{N}(\mathbf{0}, \mathbf{A}\mathbf{C}_{k,l}\mathbf{A}^T + \sigma_m^2 \mathbf{I}_N).
\end{equation}
The noisy measurements with pdf \eqref{eq:pdf y} are quantized such that
\begin{equation} \label{eq:VQ}
    \mathbf{y_q} = \mathcal{Q}(\mathbf{Ax} + \mathbf{n_m}) = \mathbf{y} + \mathbf{n_q},
\end{equation}
where~$\mathbf{y_q} \in \mathbb{R}^{N}$ is the quantized noisy CS measurement vector, and~$\mathbf{n_q} \in \mathbb{R}^{N}$ is quantization noise. Here, we model the quantization error as an additive random variable. Such structure can be found for example in \cite{96:Widrow}. Note that since the sampling rate for sparse signals is relatively low, it is reasonable to employ high-resolution quantizers; thus, as $\mathbf{y}$ has a GM density, we exploit high-rate GMM-VQ~\cite{00:Hedelin,03:Subramaniam,08:Saikat1,08:Saikat2}. Although efforts have been put into deriving the asymptotic distribution of the quantization noise (see e.g., \cite{96:Lee,00:Hedelin}), its distribution does not precisely follow a common analytic expression. Therefore, we require to derive statistical moments of the overall noise according to the VQ and Gaussian noise parameters. The $\eta$-moment quantization distortion for the GM measurements is defined as
\begin{equation} \label{eq:dist_y}
    \Delta_{\eta} \triangleq \mathbb{E}[\|\mathbf{y} - \mathbf{y_q} \|_2^\eta] = \mathbb{E}[\|\mathbf{n_q}\|_2^\eta].
\end{equation}
For brevity, we simply say quantization distortion for $\Delta_{\eta\!=\!2}$. In order to find $\Delta_{\eta}$ for the GMM-VQ, we take a linearized approach~\cite{03:Subramaniam} which characterizes the overall distortion as the sum of weighted distortion for each sub-vector, i.e.,
\begin{equation} \label{eq:linear dist}
    \Delta_{\eta} \approx \sum_{k=1}^K \sum_{l=1}^{L_k} \omega_{k,l} \Delta_{\eta,k,l} (b_{k,l}),
\end{equation}
where $b_{k,l}$ and $\Delta_{\eta,k,l}$ are the allocated bits and incurred distortion of the VQ designed for the $(k,l)^{th}$ Gaussian component, respectively. Assuming the Gaussian components in \eqref{eq:pdf y} are relatively far apart, i.e., in the limit case of \defref{def1}, then using high-rate VQ analysis for a GM source, $\Delta_{\eta,k,l} (b_{k,l})$ is given by~\cite{90:Gray}
\begin{equation} \label{eq:inter cluster dist}
    \Delta_{\eta,k,l} (b_{k,l}) \approx \left(2^{b_{k,l}} \right)^{-\frac{\eta}{N}} V_{\eta,N} |\mathbf{AC}_{k,l}\mathbf{A}^T + \sigma_m^2 \mathbf{I}_N|^{\frac{\eta}{2N}},
\end{equation}
$\forall 1\leq k \leq K,1\leq l\leq L_k $. Moreover, $V_{\eta,N}$ is the dimensionality dependent constant defined as
\begin{equation} \label{eq:V}
V_{\eta,N}=(\sqrt{2})^\eta \left(\frac{N}{2} \Gamma\left(\frac{N}{2}\right) \right)^{\frac{\eta}{N}} \left(\frac{N+\eta}{N}\right)^{\frac{N+\eta-2}{2}},
\end{equation}
where $\Gamma(\cdot)$ is the Gamma function. For the sake of theory, we assume \eqref{eq:inter cluster dist} holds with equality where practical experiments show that this assumption is precise \cite[Chapter 7]{08:Saikat_PHD}.

The quantized measurements are then sent over an AWGN channel that results $\mathbf{y_c}\!=\! \mathbf{Ax} \!+\! \mathbf{n_m} \!+\! \mathbf{n_q} \!+ \! \mathbf{n_c} \! \triangleq \! \mathbf{Ax} \!+\! \mathbf{n} $,
in which $\mathbf{n_c} \in \mathbb{R}^N$ is independent of $\mathbf{n_q}$ and $\mathbf{n_m}$, and is distributed as $\mathcal{N}\left(\mathbf{0},\sigma_c^2 \mathbf{I}_N\right)$ where $\sigma_c^2$ is known \textit{a priori}. Furthermore, $\mathbf{y_c}$ is the noisy channel output which is fed to the reconstruction algorithm. Further, it can be shown that 
\begin{equation} \label{eq:noise moments}
    \mathbb{E}[\|\mathbf{n}\|_2^2] \!=\! \mathbb{E}[\|\mathbf{n_q}\!+\! \mathbf{n_m} \!+\! \mathbf{n_c}\|_2^2] \!=\! \Delta_{\eta=2} \!\!+ \!\!N (\sigma_m^2 \!+\! \sigma_c^2),
\end{equation}
\begin{equation} \label{eq:noise moments2}
\begin{aligned}
     &\textrm{Var} [\|\mathbf{n}\|_2^2] \!\!=\!\! \textrm{Var} [\|\mathbf{n_q}\|_2^2] \!+\! 4N(\sigma_m^2 \!+\! \sigma_c^2) \mathbb{E}[\|\mathbf{n_q}\|_2^2] \!+\! 2N (\sigma_m^2 \!+ \! \sigma_c^2)^2 &\\
    &\!=\! \Delta_{\eta=4} \!-\! \Delta_{\eta=2}^2 \!+\! 4N (\sigma_m^2 \!+\! \sigma_c^2) \Delta_{\eta=2} \!+\! 2N(\sigma_m^2 \!+\! \sigma_c^2)^2 .&
\end{aligned}
\end{equation}
\vspace{-0.2cm}
\subsection{Reconstruction Algorithms} \label{reconstruction algorithm}
\vspace{-0.1cm}
For the purpose of sparse reconstruction, we first consider a generalization of the \textit{Basis pursuit de-nosing} (BPDN) algorithm for recovery of a block sparse signal which reads as follows.
\begin{theo} \cite[Theorem 2]{09:Eldar}  \label{theo0}
Let $\mathbf{x}$ be a $K$ block sparse vector, and let $\mathbf{x}_r^K$ denote the best block $K$-sparse approximation of $\mathbf{x}_r$ in the set $\mathcal{R}$. Given $\mathbf{A}$ holds the block-RIP \eqref{eq:BRIP} with constant $\delta_{2K|\mathcal{R}} < \sqrt{2}-1$, the block BPDN program
    \begin{equation} \label{eq:bpdn}
        \mathbf{\hat{x}_{BP}} = \underset{\mathbf{x}}{\textrm{argmin}} \sum_{r=1}^R \|\mathbf{x}_r \|_2 \hspace{0.25cm} \textrm{s.t.} \hspace{0.25cm} \|\mathbf{y_c} - \mathbf{Ax} \|_2 \leq \epsilon
    \end{equation}
guarantees that $\|\mathbf{x}- \mathbf{\hat{x}_{BP}} \|_2 \leq$
\begin{equation} \label{eq:guarantee}
    \frac{4\epsilon \sqrt{1\!+\!\delta_{2K|\mathcal{R}}}}{1\!-\!(1\!+\!\sqrt{2})\delta_{2K|\mathcal{R}}} +\frac{2(1\!-\!\delta_{2K|\mathcal{R}})}{1-(1+\sqrt{2})\delta_{2K|\mathcal{R}}}K^{-1/2} \sum_{r=1}^R\|\mathbf{x}_r \!-\! \mathbf{x}_r^K\|_2.
\end{equation}
\end{theo}

We assume that $\mathbf{x}$ follows \defref{def1} in the limit case, therefore, the second term in the upper-bound \eqref{eq:guarantee} vanishes. Thus, the stability condition \eqref{eq:guarantee} relies upon $\|\mathbf{n}\|_2$ and block-RIP constant $\delta_{2K|\mathcal{R}}$ which depend on the VQ and noise parameters, and on the block sparsity and $\mathbf{A}$, respectively. 

Using all practical reconstruction algorithms such as \eqref{eq:bpdn}, one should always pay the penalty that non-zero coefficient locations (the support set) are not perfectly found. However, if such algorithm exists that completely identifies the support set, the linear least-square is the best estimator in this case which is usually referred to as oracle-assisted estimator. Hence, the reconstruction error using all practical recovery algorithms, such as BPDN, is always lower-bounded by that of the oracle-assisted estimator.

\section{Main Results} \label{section:analysis}
\vspace{-0.2cm}
We quantify the system performance by estimation error of the reconstruction algorithms, i.e., $\|\mathbf{x}\!-\!\mathbf{\hat{x}} \|_2$. For this purpose, we first optimally design the rate allocations of the VQ by minimizing the quantization distortion. The following lemma proved in Appendix gives the optimal rate allocations.
\begin{lem} \label{lem:optimal}
The optimal bit allocations that minimize $\Delta_{\eta\!=\!2}$ in \eqref{eq:linear dist} under total quantization rate $b_{t}$ is given by
\begin{equation} \label{eq:bit allocation}
    2^{b_{k,l}^\star} = 2^{b_{t}} \frac{\left[\omega_{k,l}  |\mathbf{AC}_{k,l}\mathbf{A}^T+ \sigma_m^2 \mathbf{I}_N|^{\frac{1}{N}}\right]^{\frac{N}{N+2}}}{\sum_k \sum_l \left[\omega_{k,l} |\mathbf{AC}_{k,l}\mathbf{A}^T+ \sigma_m^2 \mathbf{I}_N|^{\frac{1}{N}}\right]^{\frac{N}{N+2}}},
\end{equation}
which gives the minimum quantization distortion $\Delta_{\eta=2}^{\star} =$
\begin{equation} \label{eq:first moment}
     \left(2^{b_{t}}\right)^{-\frac{2}{N}} V_{2,N} \left[\sum_{k} \sum_l \left[\omega_{k,l} |\mathbf{AC}_{k,l}\mathbf{A}^T+ \sigma_m^2 \mathbf{I}_N|^{\frac{1}{N}} \right]^{\frac{N}{N+2}} \right]^{\frac{N+2}{N}}.
\end{equation}
where $V_{2,N}$ is computed by~\eqref{eq:V} by setting $\eta=2$
\end{lem}

The recovery guarantee using the block-BPDN is given by the following theorem proved in Appendix.
\begin{theo} \label{theo1}
Let $\mathbf{x}$ be a $K$-block sparse vector that follows \defref{def1}, and let $\mathbf{A}$ hold the block-RIP \eqref{eq:BRIP} with constant $\delta_{2K|\mathcal{R}} \!<\! \sqrt{2}-1$. Given the distortion-minimizing GMM-VQ with total quantization rate $b_{t}$ and the block BPDN \eqref{eq:bpdn}, then for some $a>0$ the estimation error is upper-bounded by
\begin{equation} \label{eq:bound bpdn}
\vspace{-0.1cm}
    \|\mathbf{x} \!-\! \mathbf{\hat{x}_{BP}} \|_2 \leq \frac{4\sqrt{1+\delta_{2K|\mathcal{R}}}}{1-(1+\sqrt{2})\delta_{2K|\mathcal{R}}} \sqrt{ a + \beta + N (\sigma_m^2+\sigma_c^2)},
\vspace{-0.1cm}
\end{equation}
with probability of exceeding
\begin{equation} \label{eq:prob bound}
\vspace{-0.1cm}
1 \!-\! \frac{\alpha - \beta^2 + 4N(\sigma_m^2+\sigma_c^2) \beta + 2N(\sigma_m^2 + \sigma_c^2)^2}{a^2 + \alpha - \beta^2 +  4N(\sigma_m^2+\sigma_c^2) \beta + 2N(\sigma_m^2 + \sigma_c^2)^2},
\vspace{-0.1cm}
\end{equation}
where $\alpha= \left(2^{b_{t}}\right)^{-\frac{4}{N}} V_{4,N} \cdot$
\begin{equation} \label{eq:alpha}
  \vspace{-0.3cm}
   \frac{\sum_{k,l} \left[\omega_{k,l} |\mathbf{AC}_{k,l}\mathbf{A}^T \!+\! \sigma_m^2 \mathbf{I}_N|^{\frac{N\!-\!6}{2N(N\!-\!2)}} \right]^{\frac{N\!-\!2}{N\!+\!2}}}{\left[\sum_{m,n}  \left[\omega_{m,n} |\mathbf{AC}_{m,n}\mathbf{A}^T + \sigma_m^2 \mathbf{I}_N|^{\frac{1}{N}} \right]^{\frac{N}{N+2}} \right]^{\frac{-4}{N}}},
\vspace{-0.1cm}
\end{equation}
\begin{equation} \label{eq:beta}
  \beta \!\!=\!\! \left(2^{b_{t}}\right)^{-\frac{2}{N}} V_{2,N} \left[\sum_{k,l}  \left[\omega_{k,l} |\mathbf{AC}_{k,l}\mathbf{A}^T + \sigma_m^2 \mathbf{I}_N|^{\frac{1}{N}} \right]^{\frac{N}{N+2}} \right]^{\frac{N+2}{N}}.
\end{equation}
\end{theo}
\vspace{-0.4cm}
The following corollary is proved in Appendix.
\begin{coro} \label{coro1}
Let $\mathbf{x}$ be a $K$-block sparse vector that follows \defref{def1}, and let $\mathbf{A}$ hold the block-RIP \eqref{eq:BRIP} with constant $\delta_{K|\mathcal{R}}$. Given the distortion-minimizing GMM-VQ with total quantization rate $b_{t}$ and an oracle-assisted estimator, then for some $a>0$ the estimation error is lower-bounded as
\vspace{-0.3cm}
\begin{equation} \label{eq:bound bpdn}
    \|\mathbf{x} - \mathbf{\hat{x}_{or}} \|_2 \geq  \sqrt{a + \beta + N(\sigma_m^2 + \sigma_c^2)}/\sqrt(1+\delta_{K|R}),
    \vspace{-0.3cm}
\end{equation}
with probability of not exceeding the complement of \eqref{eq:prob bound}.
\end{coro}

\vspace{-0.2cm}
\section{Simulation results} \label{sec:numerical}
\vspace{-0.1cm}
It is common to measure the performance using signal-to-reconstruction-noise ratio (SRNR) which is defined as $\textrm{SRNR} \! \triangleq \! \mathbb{E}[\|\mathbf{x}\|_2^2]/\|\mathbf{x}-\mathbf{\hat{x}}\|_2^2$,
where $\mathbf{\hat{x}}$ is the reconstructed signal vector. Note that the upper- (lower-) bounds  on the estimation error are equivalent to lower- (upper-) bounds on SRNR. In addition, the higher SRNR achieves, the better the performance is. We choose the dimension of the source, number of blocks and block sparsity level as $M\!=\!300$, $R\!=\!30$ and $K\!=\!1$, respectively. Therefore, this level of block sparsity is equivalent to $3.33\%$ sparsity level in the conventional sense. In all simulation cases, we have assumed equal probabilities for different arrangements of the information-bearing block in the vector $\mathbf{x}$, therefore, $\omega_{k,l}\!=\! 1/R$. We randomly generate a set of 1-block sparse data $\mathbf{x}$ where the support set is chosen uniformly over the set $\{1,\ldots,R\}$. In one realization of the vector $\mathbf{x}$, the entries of the dominant sub-vector are independently drawn from a zero-mean Gaussian source with variance $\rho_x^2 \!=\! 1$, and the components of the remaining sub-vectors are chosen independently from a zero-mean Gaussian source with variance $\theta^2 \!=\! 10^{-10}$. The sensing matrix $\mathbf{A}$ is randomly generated where the components are drawn from a Gaussian source, i.e., $a_{i,j} \!\sim \!\mathcal{N} (0,\frac{1}{N})$, and then scale the columns to unit norm. We assume the measurement and channel noises are negligible. Now, let us define the fraction of measurements (FoM) as $\textrm{FoM} \!\triangleq \!N/M$ varying form 0.05 to 1 in a step size of $0.05$.

In our experiment, we compute the median of SRNR using the reconstruction algorithms which is an appropriate criterion of typical estimation error. In other words, the performance bounds are guaranteed with a $50\%$ confidence interval. Therefore, fixing the success probability \eqref{eq:prob bound} to $0.5$, the constant $a$ can be determined. We have computed $\delta_{2|\mathcal{R}}$ using Brute-force search. For this purpose, we have generated a 2-block sparse data $\mathbf{x}$ in the same way as described earlier, and find a maximum value of $\delta_{2|\mathcal{R}}$ that satisfies the lower-bound in \eqref{eq:BRIP}. This gives $\delta_{2|\mathcal{R}}\!<\!\sqrt{2}-1$ for $\textrm{FoM}\! \geq \!0.5$ for which the upper-bound in \theoref{theo1} using the block-BPDN is valid. Similarly, $\delta_{1|\mathcal{R}}$ has been derived for the oracle bound.

\begin{figure}
  \begin{center}
  \includegraphics[width=1\linewidth,height=5.8cm]{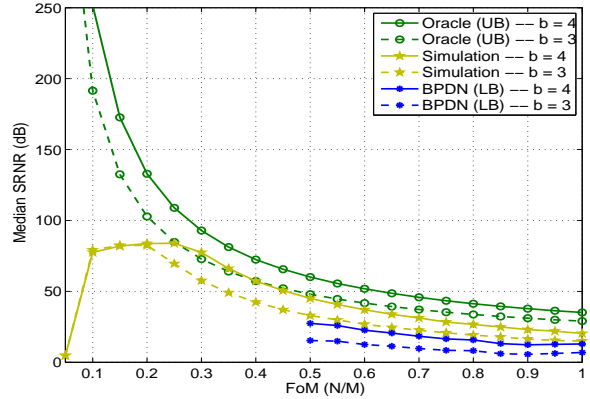}\\
  \caption{Median SRNR as a function of FoM for different quantization rates.} 
  \label{fig:result1}
  \end{center}
  \vspace{-0.9cm}
\end{figure}

\figref{fig:result1} shows the lower-bound (\theoref{theo1}) and the upper-bound (\cororef{coro1}) on SRNR in terms of FoM for various amounts of total quantization rates $b\!=\!b_t/M$ bits/scalar using the block-BPDN and the oracle-assisted algorithms. The theoretical bounds show that at a fixed $b$, as FoM increases, the lower-bounds on SRNR will decrease which is due to the fact that the quantization noise level increases with number of measurements $N$. Increasing $b$ improves the performance since it decreases the quantization noise level. To observe the effect of quantization noise in a practical simulation setup, we plot the median of actual SRNR. For this purpose, we first generate quantization noise based on the following model. It is well-known that the quantization noise for each measurement approximately behaves like a uniform RV on $(-q/2,q/2)$ at high rates, where $q\!=\!1/(2^{Mb/N-1})$ is the distance between quantization levels. Under this model, the probability that $\|\mathbf{n_q}\|_2$ exceeds the added value of the mean and two/three times the standard deviation is small. Therefore, as in \cite{06:Candes2} we choose $\epsilon\!=\!\sqrt{Nq^2/12 \!+\! 3\sqrt{N}q^2/(6\sqrt{5})}$, and solve \eqref{eq:bpdn} using a standard convex solver for 100 realizations of $\mathbf{x}$ for each value of FoM. Finally, among all sorted values of SRNR, we pick the middle ones which give the medians.

In \figref{fig:result1}, it can be seen that both number of measurements $N$ and total quantization rate $b$ influence the performance of actual curves. These impacts can be interpreted as follows. Given a very small FoM, the number of measurements are not enough in order for the block-BPDN algorithm to reconstruct the source. As FoM increases to a certain amount, the algorithm is able to recover the block sparse source precisely out of the measurements since the number of measurements are sufficient and the quantization noise level is small enough. However, for a higher FoM, due to the limited rate, the quantization noise per component increases which leads to the poor performance. 
The intuition behind the behavior of the oracle bounds, as compared to the actual curves, is that the oracle-assisted estimator has the knowledge of the support set (i.e., location of dominant entries of $\mathbf{x}$) even from small number of observations, unlike practical sparse recovery algorithms. It is also worth to recall that the estimation error of any practical reconstruction method is always larger than that of the oracle-assisted algorithm, thus, the oracle bound casts as a standard upper-bound on SRNR of the block-BDPN scheme.

\vspace{-0.25cm}
\section{Conclusion} \label{sec:conclusion}
\vspace{-0.15cm}
We have addressed the design of VQ for block sparse signals using block sparse recovery algorithms. Inspired by a GMM for block sparse sources, optimal rate allocation has been designed for a GMM-VQ which aims to minimize quantization distortion. We have theoretically derived upper- and lower-bounds on estimation error of block BPDN algorithm and oracle-assisted estimator, respectively, by taking the optimal VQ parameters into account in the presence of AWGN.
\vspace{-0.2cm}
\appendix \label{app1}
\vspace{-0.15cm}
\noindent \textit{\textbf{Proof of \lemref{lem:optimal}:}}
we minimize $\Delta_{\eta\!=\!2}$ under the total quantization rate available for the VQ, i.e., we solve
\begin{equation*}
        \underset{\{b_{k,l}\}}{\textrm{min}} \hspace{0.4cm} \sum_{k=1}^{K} \sum_{l=1}^{L_k} \omega_{k,l} \left(2^{b_{k,l}} \right)^{-\frac{2}{N}} V_{2,N} |\mathbf{AC}_{k,l}\mathbf{A}^T + \sigma_m^2 \mathbf{I}_N|^{\frac{1}{N}}
\end{equation*}
\begin{equation} \label{eq:opt problem}
        \textrm{\hspace{-3cm} subject to}  \hspace{0.5cm}   \sum_{k=1}^{K} \sum_{l=1}^{L_k} 2^{b_{k,l}} = 2^{b_{t}}.
\end{equation}
Solving \eqref{eq:opt problem} is based on the fact that the optimization problem is convex in $b_{k,l}$. Therefore, using the Lagrangian method and Karush-–Kuhn–-Tucker (KKT) conditions, the optimal $b_{k,l}$ and $\Delta_{\eta=2}$ can be obtained explicitly.

\noindent \textit{\textbf{Proof of \theoref{theo1}:}}

We show that for an appropriate choice of $\epsilon$ in \eqref{eq:bpdn}, the estimation error~$\|\mathbf{x}\!-\!\mathbf{\hat{x}} \|_2$ is bounded with overwhelming probability. We proceed with the Cantelli inequality, i.e., for any random variable $\xi$ and a constant $a > 0$,
\begin{equation} \label{eq:Cantelli}
    \Pr\{\xi \geq a + \mathbb{E}[\xi] \} \leq \frac{\mathrm{Var}[\xi]}{a^2+\mathrm{Var}[\xi]}.
\end{equation}
It is reasonable to employ the concentration inequality~\eqref{eq:Cantelli} since it takes both the first and second moments of the unknown RV $\xi$ into account. Now, let $C=4 \sqrt{1\!+\!\delta_{2K|\mathcal{R}}}/(1\!-\!(1\!+\!\sqrt{2})\delta_{2K|\mathcal{R}})$, then for some $a>0$, we have
\begin{equation}
\begin{aligned}
&\Pr\left\{\|\mathbf{x}\!-\!\mathbf{\hat{x}_{BP}}\|_2^2 \!\geq\! C^2\left(a \!+\! \Delta_{\eta\!=\!2}^{\star}\!+\!\mathbb{E}[\|\mathbf{n_m + n_c}\|_2^2]\right)\right\}& \\
&\! \stackrel{(a)}{\leq} \! \Pr \left\{\|\mathbf{n}\|_2^2 \!\geq\! a \!+\! \Delta_{\eta\!=\!2}^{\star}\!+\!\mathbb{E}[\|\mathbf{n_m + n_c}\|_2^2]\right\}
\stackrel{(b)}{\leq} \frac{\mathrm{Var}[\|\mathbf{n}\|_2^2]}{a^2 \!+\! \mathrm{Var}[\|\mathbf{n}\|_2^2]},&
\end{aligned}
\end{equation}
where $(a)$ follows from the stability guarantee \eqref{eq:guarantee} by considering only the first term, and $(b)$ from \eqref{eq:Cantelli} by substituting $\xi$ with $\|\mathbf{n}\|_2^2$. Finally, using optimal bit allocations and combining \eqref{eq:noise moments}, \eqref{eq:noise moments2} and \eqref{eq:first moment}, the proof completes.

\noindent \textit{\textbf{Proof of \cororef{coro1}:}}
To provide a lower-bound on the estimation error using the oracle-assisted estimator, we first denote the support set of $\mathbf{x}$ by $\Omega\! \triangleq \!\{i\in \{1,\ldots,R\} :\|\mathbf{x}_i\|_2 > \gamma\}$, for a small positive $\gamma$, with block sparsity \textbf{card}$\{\Omega\}=K$. If the support set is perfectly reconstructed, the oracle estimate satisfies $\mathbf{\hat{x}}|_{\Omega} \!=\! \mathbf{x}|_\Omega \!+\! \mathbf{A}_\Omega^\dag \mathbf{n}$, where $\mathbf{\hat{x}}|_{\Omega}$ denotes the reconstructed signal in the support set $\Omega$, $\mathbf{A}_\Omega$ denotes the sub-matrix of $\mathbf{A}$ formed by choosing the columns of $\mathbf{A}$ indexed by $\Omega$, and $\dag$ is the pseudo inverse. Further, $\mathbf{x}|_\Omega$ possesses the entries of $\mathbf{x}$ indexed by $\Omega$. Therefore, for some $a>0$ we have
\begin{equation} \label{proof2}
\begin{aligned}
&\Pr\left\{\|\mathbf{x} \!-\! \mathbf{\hat{x}_{or}} \|_2^2 \!\leq\! \frac{a \!+\! \Delta_{\eta\!=\!2}^{\star}\!+\!\mathbb{E}[\|\mathbf{n_m + n_c}\|_2^2]}{1\!+\!\delta_{K|R}}\right\}&\\
&\!=\! \Pr\left\{\|\mathbf{A}_\Omega^\dag \mathbf{n}\|_2^2 \!\leq \! \frac{a \!+\! \Delta_{\eta\!=\!2}^{\star}\!+\!\mathbb{E}[\|\mathbf{n_m + n_c}\|_2^2]}{1\!+\!\delta_{K|R}}\right\}& \\
&\stackrel{(a)}{\geq} \Pr \left\{s_{\min}^2(\mathbf{A}_\Omega^\dag) \|\mathbf{n}\|_2^2 \!\leq \! \frac{a \!+\! \Delta_{\eta\!=\!2}^{\star}\!+\!\mathbb{E}[\|\mathbf{n_m + n_c}\|_2^2]}{1\!+\!\delta_{K|R}} \right\}& \\
&\stackrel{(b)}{\geq} \Pr\left\{\|\mathbf{n}\|_2^2 \leq a + \mathbb{E}[\|\mathbf{n}\|_2^2] \right\}
\stackrel{(c)}{\geq} 1 - \frac{\mathrm{Var}[\|\mathbf{n}\|_2^2]}{a^2 + \mathrm{Var}[\|\mathbf{n}\|_2^2]},&
\end{aligned}
\end{equation}
where in $(a)$, $s_{\min}$ denotes the minimum singular-value, and we use the inequality $s_{\min}^2(\mathbf{A}_\Omega^\dag)\|\mathbf{n}\|_2^2 \! \leq \! \|\mathbf{A}_\Omega^\dag \mathbf{n}\|_2^2$. $(b)$ follows from the fact that the minimum singular value of $\mathbf{A}_\Omega^\dag$ is lower-bounded by $1/\sqrt{1+\delta_{K|R}}$ \cite{11:Laska}. $(c)$ follows by \eqref{eq:Cantelli}, and by combining it with \eqref{eq:noise moments}, \eqref{eq:noise moments2} and \eqref{eq:first moment} the proof completes.

\vspace{-0.2cm}

\bibliographystyle{IEEEtran}
\bibliography{IEEEfull,bibliokthPasha}

\begin{thebibliography}{10}
\providecommand{\url}[1]{#1}
\csname url@samestyle\endcsname
\providecommand{\newblock}{\relax}
\providecommand{\bibinfo}[2]{#2}
\providecommand{\BIBentrySTDinterwordspacing}{\spaceskip=0pt\relax}
\providecommand{\BIBentryALTinterwordstretchfactor}{4}
\providecommand{\BIBentryALTinterwordspacing}{\spaceskip=\fontdimen2\font plus
\BIBentryALTinterwordstretchfactor\fontdimen3\font minus
  \fontdimen4\font\relax}
\providecommand{\BIBforeignlanguage}[2]{{%
\expandafter\ifx\csname l@#1\endcsname\relax
\typeout{** WARNING: IEEEtran.bst: No hyphenation pattern has been}%
\typeout{** loaded for the language `#1'. Using the pattern for}%
\typeout{** the default language instead.}%
\else
\language=\csname l@#1\endcsname
\fi
#2}}
\providecommand{\BIBdecl}{\relax}
\BIBdecl

\bibitem{08:Candes}
E.~Candes and M.~Wakin, ``An introduction to compressive sampling,'' \emph{IEEE
  Signal Processing Magazine}, vol.~25, no.~2, pp. 21 --30, Mar. 2008.

\bibitem{06:Candes2}
E.~Candes, J.~Romberg, and T.~Tao, ``Stable signal recovery from incomplete and
  inaccurate measurements,'' \emph{Comm. Pure Appl. Math}, vol.~59, no.~8, pp.
  1207--1223, 2006.

\bibitem{07:Tropp}
J.~Tropp and A.~Gilbert, ``Signal recovery from random measurements via
  orthogonal matching pursuit,'' \emph{IEEE Trans. Inf. Theory}, vol.~53,
  no.~12, pp. 4655 --4666, Dec. 2007.

\bibitem{07:Candes}
E.~Candes and T.~Tao, ``Rejoinder: the {Dantzig} selector: statistical
  estimation when $p$ is much larger than $n$,'' \emph{Annals of Statistics},
  vol.~35, pp. 2392 -- 2404, 2007.

\bibitem{08:Shihao}
S.~Ji, Y.~Xue, and L.~Carin, ``Bayesian compressive sensing,'' \emph{IEEE
  Trans. Sig. Proc.}, vol.~56, no.~6, pp. 2346 --2356, Jun. 2008.

\bibitem{09:Dai}
W.~Dai and O.~Milenkovic, ``Subspace pursuit for compressive sensing signal
  reconstruction,'' \emph{IEEE Trans. Inf. Theory}, vol.~55, no.~5, pp. 2230
  --2249, May 2009.

\bibitem{10:Zymnis}
A.~Zymnis, S.~Boyd, and E.~Candes, ``Compressed sensing with quantized
  measurements,'' \emph{IEEE Signal Processing Letters}, vol.~17, no.~2, pp.
  149 --152, Feb. 2010.

\bibitem{11:Dai}
W.~Dai and O.~Milenkovic, ``Information theoretical and algorithmic approaches
  to quantized compressive sensing,'' \emph{IEEE Trans. on Communications},
  vol.~59, no.~7, pp. 1857 --1866, Jul. 2011.

\bibitem{11:Laska}
J.~N. Laska and R.~G. Baraniuk, ``Regime change: Bit-depth versus
  measurement-rate in compressive sensing,'' \emph{CoRR}, vol. abs/1110.3450,
  2011.

\bibitem{11:Jacques}
L.~Jacques, D.~Hammond, and J.~Fadili, ``Dequantizing compressed sensing: When
  oversampling and non-{Gaussian} constraints combine,'' \emph{IEEE Trans. Inf.
  Theory}, vol.~57, no.~1, pp. 559 --571, Jan. 2011.

\bibitem{09:Stojnic}
M.~Stojnic, F.~Parvaresh, and B.~Hassibi, ``On the reconstruction of
  block-sparse signals with an optimal number of measurements,'' \emph{IEEE
  Trans. Sig. Proc.}, vol.~57, no.~8, pp. 3075 --3085, Aug. 2009.

\bibitem{10:Stojnic}
M.~Stojnic, ``$l_{2}/l_{1}$--optimization in block-sparse compressed sensing
  and its strong thresholds,'' \emph{IEEE Journal of Selected Topics in Signal
  Processing}, vol.~4, no.~2, pp. 350 --357, Apr. 2010.

\bibitem{10:Eldar}
Y.~Eldar, P.~Kuppinger, and H.~B\"{o}lcskei, ``Block-sparse signals:
  {Uncertainty} relations and efficient recovery,'' \emph{IEEE Trans. Sig.
  Proc.}, vol.~58, no.~6, pp. 3042 --3054, Jun. 2010.

\bibitem{10:Baraniuk}
R.~Baraniuk, V.~Cevher, M.~Duarte, and C.~Hegde, ``Model-based compressive
  sensing,'' \emph{IEEE Trans. Inf. Theory}, vol.~56, no.~4, pp. 1982 --2001,
  Apr. 2010.

\bibitem{00:Hedelin}
P.~Hedelin and J.~Skoglund, ``Vector quantization based on {Gaussian} mixture
  models,'' \emph{IEEE Trans. on Speech and Audio Processing}, vol.~8, no.~4,
  pp. 385 --401, Jul. 2000.

\bibitem{03:Subramaniam}
A.~Subramaniam and B.~Rao, ``{PDF} optimized parametric vector quantization of
  speech line spectral frequencies,'' \emph{IEEE Trans. on Speech and Audio
  Processing}, vol.~11, no.~2, pp. 130 -- 142, Mar. 2003.

\bibitem{08:Saikat1}
S.~Chatterjee and T.~Sreenivas, ``Switched conditional {PDF}-based split {VQ}
  using gaussian mixture model,'' \emph{IEEE Signal Processing Letters},
  vol.~15, pp. 91 --94, Jan. 2008.

\bibitem{08:Saikat2}
------, ``Predicting {VQ} performance bound for {LSF} coding,'' \emph{IEEE
  Signal Processing Letters}, vol.~15, pp. 166 --169, Jan. 2008.

\bibitem{11:Mikko}
M.~Vehkapera, S.~Chatterjee, and M.~Skoglund, ``Analysis of {MMSE} estimation
  for compressive sensing of block sparse signals,'' in \emph{IEEE Inf. Theory
  Workshop}, Oct. 2011, pp. 553 --557.

\bibitem{09:Eldar}
Y.~Eldar and M.~Mishali, ``Robust recovery of signals from a structured union
  of subspaces,'' \emph{IEEE Trans. Inf. Theory}, vol.~55, no.~11, pp. 5302
  --5316, Nov. 2009.

\bibitem{05:Candes}
E.~Candes and T.~Tao, ``Decoding by linear programming,'' \emph{IEEE Trans.
  Inf. Theory}, vol.~51, no.~12, pp. 4203 -- 4215, Dec. 2005.

\bibitem{06:Candes}
E.~Candes, J.~Romberg, and T.~Tao, ``Robust uncertainty principles: exact
  signal reconstruction from highly incomplete frequency information,''
  \emph{IEEE Trans. Inf. Theory}, vol.~52, no.~2, pp. 489 -- 509, Feb. 2006.

\bibitem{96:Widrow}
B.~Widrow, I.~Kollar, and M.-C. Liu, ``Statistical theory of quantization,''
  \emph{IEEE Trans. Instr. Meas.}, vol.~45, no.~2, pp. 353 --361, Apr. 1996.

\bibitem{96:Lee}
D.~Lee and D.~Neuhoff, ``Asymptotic distribution of the errors in scalar and
  vector quantizers,'' \emph{IEEE Trans. Inf. Theory}, vol.~42, no.~2, pp. 446
  --460, Mar. 1996.

\bibitem{90:Gray}
R.~M. Gray, \emph{Source Coding Theory}.\hskip 1em plus 0.5em minus 0.4em\relax
  Kluwer Academic Publishers, 1990.

\bibitem{08:Saikat_PHD}
S.~Chatterjee, ``Rate-distortion performance and complexity optimizaed
  structured vector quantization,'' Ph.D. dissertation, Indian Institute of
  Science, 2008.

\end{thebibliography}
\end{document}